\documentclass[twocolumn,showpacs,preprintnumbers,amsmath,amssymb,aps,prb,longbibliography]{revtex4-2}

\usepackage{bm}
\usepackage{bbm}
\usepackage{graphicx}
\usepackage{braket}
\usepackage{epstopdf}
\usepackage{color}
\usepackage{subfigure}
\usepackage[dvipsnames]{xcolor} 
\usepackage{amsmath} 
\usepackage{amssymb} 
\usepackage{ulem} 
\usepackage{mathtools}
\usepackage{placeins} 
\usepackage{sidecap}
\usepackage{hyperref}

\def\emph{\textit}

\renewcommand{\thefootnote}{\fnsymbol{footnote}} 

\begin{document}



\title{Spin-Flip Raman Scattering on Electrons and Holes in Two-Dimensional (PEA)$_2$PbI$_4$ Perovskites}

\author{C.~Harkort,$^{1}$ D.~Kudlacik,$^{1}$ N.~E.~Kopteva,$^{1}$ D.~R.~Yakovlev,$^{1}$ M.~Karzel,$^{1}$ E.~Kirstein,$^{1}$ O.~Hordiichuk,$^{2,3}$ M.~Kovalenko,$^{2,3}$ and M.~Bayer$^{1}$} 

\affiliation{$^{1}$Experimentelle Physik 2, Technische Universit\"{a}t Dortmund, D-44221 Dortmund, Germany}
\affiliation{$^{2}$ Laboratory of Inorganic Chemistry, Department of Chemistry and Applied Biosciences,  ETH Z\"{u}rich, CH-8093 Z\"{u}rich, Switzerland}
\affiliation{$^{3}$ Laboratory for Thin Films and Photovoltaics, Empa-Swiss Federal Laboratories for Materials Science and Technology, CH-8600 D\"{u}bendorf, Switzerland}

\date{\today}





\begin{abstract}
The class of Ruddlesden-Popper type (PEA)$_2$PbI$_4$ perovskites comprises two-dimensional (2D) structures whose optical properties are determined by excitons with a large binding energy of about 260~meV. It complements the family of other 2D semiconductor materials by having the band structure typical for lead halide perovskites, that can be considered as inverted compared to conventional III-V and II-VI semiconductors. Accordingly, novel spin phenomena can be expected for them. Spin-flip Raman scattering is used here to measure the Zeeman splitting of electrons and holes in a magnetic field up to 10~T. From the recorded data, the electron and hole Land\'e factors ($g$-factors) are evaluated, their signs are determined, and their anisotropies are measured. The electron $g$-factor value changes from $+2.11$ out-of-plane to $+2.50$ in-plane, while the hole $g$-factor ranges between $-0.13$ and $-0.51$. The spin flips of the resident carriers are arranged via their interaction with photogenerated excitons. Also the double spin-flip process, where a resident electron and a resident hole interact with the same exciton, is observed showing a cumulative Raman shift. Dynamic nuclear spin polarization induced by spin-polarized holes  is detected in corresponding changes of the hole Zeeman splitting. An Overhauser field of the polarized nuclei acting on the holes as large as $0.6$~T can be achieved. 
\end{abstract} 

\maketitle

\section{Introduction}

Lead halide perovskite semiconductors attract nowadays great attention due to their remarkable potential for photonic applications~\cite{Vinattieri2021_book, Vardeny2022_book_appl, jena2019, Hansell2019}. They are available as bulk- and nanocrystals and also as two-dimensional (2D) layered materials. 2D perovskites feature exceptional optical and electrical properties. By changing the thickness of the semiconductor layers and varying the organic barriers, their band gap energy changes from the infrared up to the ultraviolet spectral range~\cite{Mao2018,blancon2020,hoye2022}. The  2D perovskites exhibit robust environmental stability~\cite{Wang2020}, which makes them promising for optoelectronic~\cite{Hansell2019,lempicka-mirek2022,Wang2020} and photovoltaic\cite{Zhang2018,Cheng2018,Sidhik2021} applications. The strong quantum confinement of electrons and holes results in excitons with large binding energies, which are additionally increased by dielectric confinement, approaching $200-500$~meV~\cite{Muljarov1995,blancon2018}. The optical properties of 2D perovskites are therefore determined by exciton absorption and emission even in ambient conditions~\cite{ishihara1989,ishihara1990}, similar to 2D semiconductors like transition metal dichalcogenides~\cite{wang2018}.

The band gap in lead halide perovskites is located at the R-point of the Brillouin zone for cubic crystal lattice and at the $\Gamma$-point for tetragonal or orthorhombic lattices~\cite{Even2015,Katan2019}. In all these cases, the states at the bottom of the conduction band and the top of the valence band have spin $1/2$. The perovskites band structure is inverted compared to conventional III-V and II-VI semiconductors, i.e., in the vicinity of the band gap the valence band is mostly formed by the s-orbitals of Pb, while the conduction band states are contributed by the p-orbitals of Pb. As a result, the spin-orbit interaction modifies mostly the valence band states (and thus the hole effective mass and $g$-factor)~\cite{Yu2016,kirstein2021,kirstein2022}, and the hyperfine interaction with the nuclear spins is much stronger for the holes than for the electrons, in contrast to conventional semiconductors~\cite{kirstein2021}. Therefore, lead halide perovskites of different dimensionalities are considered as novel model systems for spin physics, offering interesting perspectives for spintronic and quantum information applications~\cite{Vardeny2022_book_spin}.   

Recent studies show that a similar level of optical spin control can be achieved in perovskites as in conventional semiconductors. To that end, the optical and magneto-optical techniques established for studying spin-dependent phenomena were tested with respect to their suitability for lead halide perovskites: optical orientation~\cite{Giovanni2015,Nestoklon2018,Wang2018oo,Wang2019}, optical alignment~\cite{Nestoklon2018}, polarized emission in magnetic field~\cite{Zhang2015,Canneson2017,Zhang2018A}, time-resolved Faraday/Kerr rotation~\cite{odenthal2017,belykh2019,kirstein2021} and spin-flip Raman scattering~\cite{Ema2006,kirstein2022} were demonstrated. Some of them were also used to study spin properties including their dynamics in 2D perovskites. Application of high magnetic fields up to 60~T provided information on the exciton fine structure and exciton Land\'e factor ($g$-factor)~\cite{kataoka1993,hirasawa1993,tanaka2005,dyksik2020,dyksik2021,do2020b,surrente2021,Baranowski2022}. The exciton spin dynamics down to subpicosecond time scales were addressed by optical spin orientation measured by time-resolved transmission~\cite{giovanni2018,pan2020,chen2021,bourelle2020,bourelle2022}. Most of the dynamical studies were carried out above liquid nitrogen up to room temperature, where the spin relaxation times do not exceed a few picoseconds. Recently, time-resolved Kerr rotation allowed measurements of the coherent dynamics of electron spins in the (PEA)$_2$PbI$_4$ 2D perovskite~\cite{kirstein2022c}. In these experiments, longitudinal spin relaxation times up to 25~$\mu$s were found at the temperature of $1.6$~K. In addition, the electron $g$-factor was measured, showing a considerable anisotropy. At present, the experimental information on the electron and hole $g$-factors in 2D perovskites, being the key parameters for understanding and interpreting spin-dependent phenomena, is still limited, and we are also not aware of corresponding theoretical considerations.   

Spin-flip Raman scattering (SFRS) spectroscopy is another powerful magneto-optical technique in spin physics, providing direct information on the Zeeman splitting of electrons, holes, and excitons, and on the optical selection rules due to the spin level structure of exciton complexes, determined by their symmetries and exciton-carrier spin interactions~\cite{Thomas1968,Haefele1991, Sapega1994, Sirenko1997,Debus2013_CdTeQW,Debus2014_SFRS_QDs,Kudlacik2020}.  SFRS signals are strongly enhanced when the laser photon energy is tuned into resonance with the exciton. SFRS measurements are experimentally challenging due to the close spectral proximity of the spin-flip signals and the laser line, as the spin-flip Raman shift is on the order of a few hundred $\mu$eV. Recently, however, the feasibility of SFRS for measuring the electron and hole $g$-factors in CsPbBr$_3$ and MAPbI$_3$ lead halide perovskite crystals was demonstrated~\cite{kirstein2022}.

In this paper, we report on an SFRS study of the electron and hole $g$-factors in  Ruddlesden-Popper type (PEA)$_2$PbI$_4$ two-dimensional perovskites. The experiments are performed at cryogenic temperatures in magnetic fields up to 10~T, applied in different geometries in order to measure the $g$-factor anisotropy. The spin-flip signals originate from resident electrons and holes interacting with photogenerated excitons. Further, nuclear spin polarization by spin polarized holes is evidenced through corresponding shifts of the hole spin-flip line.  

\section{Experimental results}
\subsection{Optical properties of two-dimensional perovskites (PEA)$_2$PbI$_4$}

We study the 2D Ruddlesden-Popper type perovskite structure (PEA)$_2$PbI$_4$, which consists of a corner-shared network of PbI$_6$-octahedral monolayers constituting quantum wells separated by van der Waals-bonded pairs of PEA [phenethylammonium] molecules. Due to the strong quantum confinement of electrons and holes in the 2D perovskite layers, the band gap energy increases to $2.608$~eV at $T=2$~K~\cite{dyksik2020}. The reduced dimensionality and the dielectric confinement effect~\cite{Katan2019}, provided by the difference of dielectric constants between the perovskite and the PEA, strongly increase the exciton binding energy in (PEA)$_2$PbI$_4$ to $260$~meV~\cite{dyksik2020,Baranowski2022}, in comparison to  $16$~meV in bulk MAPbI$_3$~\cite{Galkowski2016}.

\begin{figure*}[hbt]
\begin{center}
\includegraphics[width = 17.8cm]{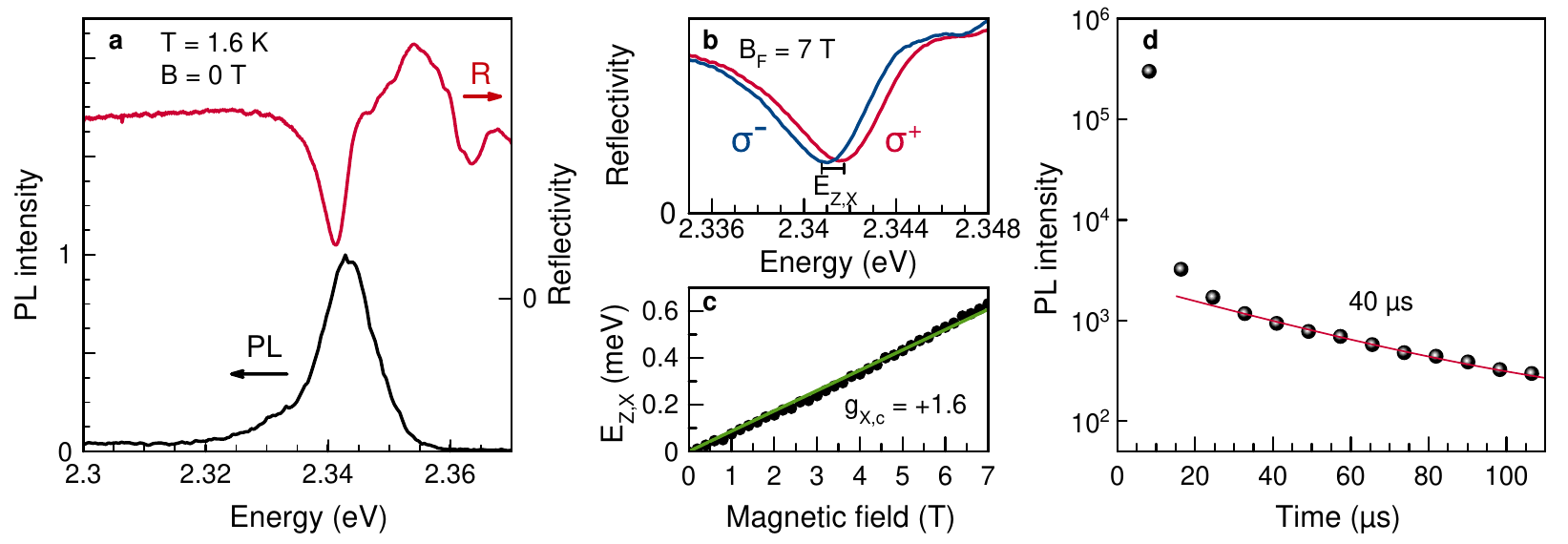}
\caption{Optical properties of excitons in (PEA)$_2$PbI$_4$ at $T=1.6$~K. 
(a) Reflectivity (red) and photoluminescence (black) spectra. The PL is excited at $2.412$~eV photon energy using $P=10.8$~W/cm$^2$ excitation power.  
(b) Counter-circularly polarized reflectivity spectra measured in Faraday geometry at $B_{\rm F}=7$~T ($\mathbf{B}_{\rm F} \parallel \mathbf{k} \parallel c$). The exciton Zeeman splitting of $E_{\rm Z,X}=0.63$~meV can be determined with good accuracy.
(c) Magnetic field dependence of the exciton Zeeman splitting evaluated from magneto-reflectivity data (symbols). The green line is a $B$-linear fit.
(d) PL dynamics (symbols) measured at the PL maximum of $2.343$~eV. Pulsed excitation is used at an energy of $3.493$~eV photon energy with an average  power of $P=3$~W/cm$^2$. The line is an exponential fit of the decay at longer time with the time constant of $40~\mu$s. 
} 
\label{fig:1}
\end{center}
\end{figure*}

The pronounced exciton resonance in (PEA)$_2$PbI$_4$ is seen in the reflectivity (R) spectrum, measured at the temperature of $T=1.6$~K, see Figure~\ref{fig:1}a. The resonance line with the minimum at 2.341~meV and the full width at half maximum of 6.6~meV originates from the free exciton. In external magnetic field applied in the Faraday geometry ($\mathbf{B}_{\rm F} \parallel \mathbf{k} \parallel c$), the exciton spin states $+1$ and $-1$ detected in $\sigma^+$ and $\sigma^-$ circular polarization, respectively, are subject to Zeeman splitting by $E_{\rm Z, X}=g_{\rm X,c}\mu_{\rm B} B_{\rm F}$. Here, the $c$-axis is perpendicular to the 2D planes, $\mathbf{k}$ is the light wave vector, $g_{\rm X,c}$ is the exciton $g$-factor along the $c$-axis, and $\mu_{\rm B}$ is the Bohr magneton. Reflectivity spectra at $B_{\rm F}=7$~T measured in $\sigma^+$ and $\sigma^-$ polarization are shown in Figure~\ref{fig:1}b. The different energies of the exciton resonance in the two spectra reflect the Zeeman splitting. In Figure~\ref{fig:1}c we present the magnetic field dependence of the exciton Zeeman splitting, from its linear fit the exciton $g$-factor $g_\text{X,c}=+1.6\pm0.1$ is evaluated. Note that in this experiment the $g$-factor sign can be determined: a positive value corresponds to a high energy shift of the $\sigma^+$ polarized resonance relative to the $\sigma^-$ polarized one.

The photoluminescence (PL) spectrum shows a strong emission line with the maximum at 2.343~eV and the full width at half maximum of about 10~meV, see Figure~\ref{fig:1}a. The PL line coincides in energy with the free exciton resonance measured in reflectivity. However, note that the PL line is broader than the reflectivity line. It is plausible to assign the line to the exciton emission of both free and weakly localized excitons. This assignment is supported by time-integrated and time-resolved spectroscopic studies at cryogenic temperatures reported in Refs.~\citenum{do2020b,do2020jcp,kahmann2021,dyksik2021,posmyk2022}, showing that the PL band is composed of at least two emission lines. Their recombination dynamics show times in the range of 300~ps to 10~ns, highlighting the free- and bound-exciton origin. 

For the studied (PEA)$_2$PbI$_4$, the population dynamics are measured by time-resolved differential reflectivity for resonant excitation of the exciton. The results are reported in Ref.~\citenum{kirstein2022c}. The dynamics trace reveals decays with times of 20~ps and 340~ps. Also a longer-lived component with a decay exceeding 1~ns is observed. We attribute the short dynamics of 20~ps to the lifetime of the bright excitons with a large oscillator strength in 2D perovskites. The lifetime is given by their radiative recombination and their relaxation into dark exciton states. This interpretation is in agreement with literature data on the low temperature recombination dynamics in (PEA)$_2$PbI$_4$~\cite{fang2020,do2020jcp,kahmann2021}. The slower 340~ps dynamics can be attributed to non-geminate recombination of charge carriers. 

In Figure~\ref{fig:1}d, the PL dynamics measured at the maximum of the PL line across a much longer temporal range up to 100~$\mu$s are shown. Recombination processes with a decay time of about 40~$\mu$s are observed, which greatly exceeds the typical times in exciton dynamics. This evidences that long-living resident carriers are present in the studied structures. These resident carriers can be photo-generated electrons and holes which are localized at spatially separated sites. We will term them as resident electrons and holes and will show that they give the main contribution to the measured SFRS signals. Note that the existence of resident carriers is typical for lead halide perovskites, as we showed for bulk CsPbBr$_3$~\cite{belykh2019}, FA$_{0.9}$Cs$_{0.1}$PbI$_{2.8}$Br$_{0.2}$~\cite{kirstein2021}, and MAPbI$_3$~\cite{kirstein2022b} crystals using optical techniques.

The weak low energy flank of the PL line at 2.330~eV, see Figure~\ref{fig:1}a, was assigned in literature either to dark exciton emission~\cite{dyksik2021,neumann2021,posmyk2022} or to phonon-assisted bright exciton recombination~\cite{do2020jcp}.

\subsection{Spin-flip Raman scattering in close-to-Faraday geometry} 
\label{sec:Faraday}

We apply spin-flip Raman scattering to study the properties of the resident carrier spins in the 2D (PEA)$_2$PbI$_4$ perovskite. In an external magnetic field, $\mathbf{B}$, the spin sublevels of the electrons (e) and the holes (h) are split by the Zeeman energy $E_{\rm Z, e(h)}=g_{\rm e(h)}\mu_{\rm B} B$, which is proportional to the magnetic field strength and the electron (hole) $g$-factor $g_{\rm e(h)}$. In the process of Raman light scattering, the carrier spin can flip changing its orientation, which requires either absorption or dissipation of the energy amount equal to $E_{\rm Z, e(h)}$, depending on whether the spin flips from the lower to the upper energy level or vice versa. Therefore, the energy of the scattered photon differs from the laser photon energy by $E_{\rm Z, e(h)}$. In case of energy absorption, the Raman shift occurs to lower energies (Stokes shift), note, however, that in SFRS experiments it is common to refer to this case as positive Raman shift. In case of energy dissipation, the shift is to larger energies (anti-Stokes shift), so that the Raman shift values are negative. For light scattering in semiconductors, an exciton serves as a mediator between light and spins~\cite{Sirenko1997,Debus2013_CdTeQW}, because the light-matter interaction is greatly enhanced at the exciton resonance.      

\begin{figure*}[hbt]
\begin{center}
\includegraphics[width=17.8cm]{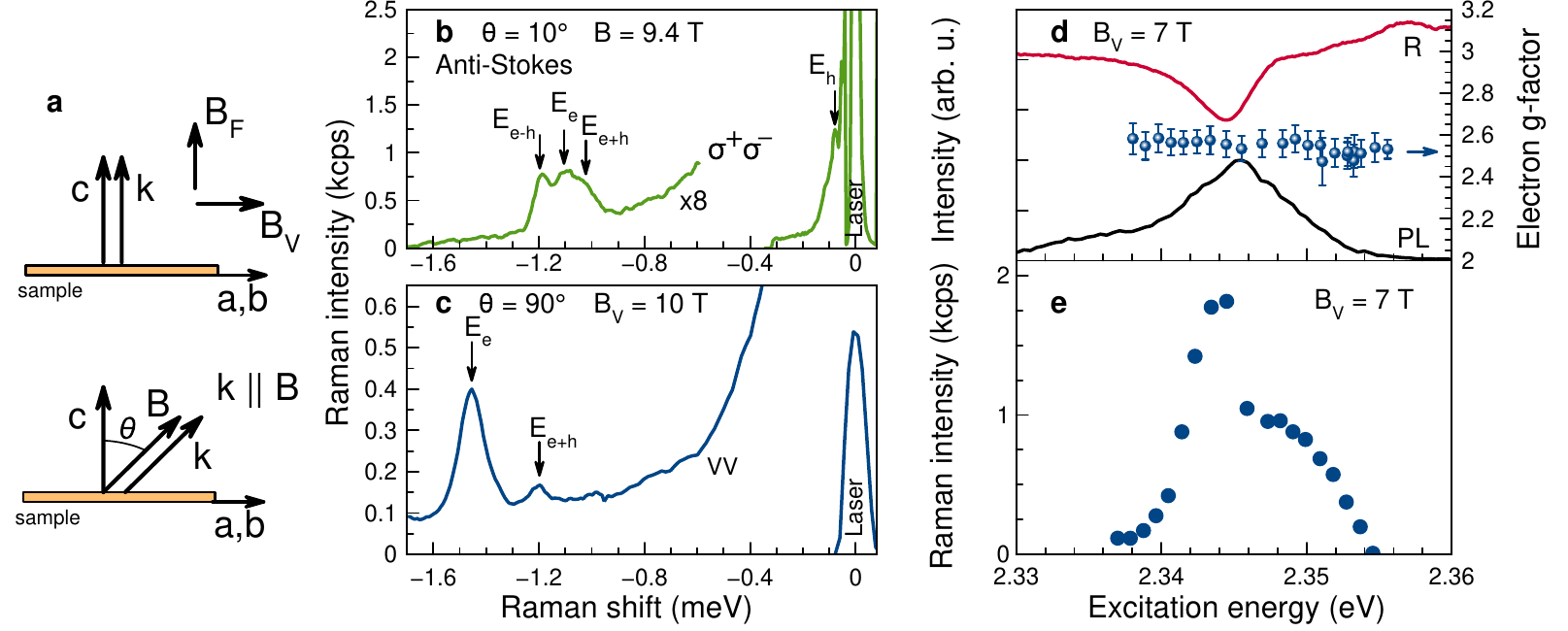}
\caption{Spin-flip Raman scattering in (PEA)$_2$PbI$_4$.
(a)  Sketch of the experimental geometry. Upper diagram is for the Faraday ($\mathbf{B}_\text{F} \parallel c$) and Voigt ($\mathbf{B}_\text{V} \parallel (a,b)$) geometry with the laser light vector $\mathbf{k} \parallel c$. Bottom diagram is for the tilted field geometry. The angle $\theta$ specifies the tilt between $\mathbf{B}$ and the $c$-axis ($\mathbf{k} \parallel \mathbf{B}$).
(b) SFRS spectrum in the anti-Stokes spectral range (negative Raman shift) for $\theta=10^{\circ}$, measured for $E_\text{exc}=2.345$~eV laser photon energy with the power $P=5.7$~W/cm$^2$. The spectrum is multiplied by a factor of 8. The shifts of  the hole $E_\text{h}$, the electron $E_\text{e}$, and their double flip ($E_\text{e+h}$ and $E_\text{e-h}$)  lines are marked with arrows. 
(c) SFRS spectrum in Voigt geometry at $B_{\rm V}=10$~T using linearly co-polarized excitation and detection.
(d) Reflectivity and photoluminescence spectra at $B_{\rm V}=7$~T. Electron $g$-factor dependence on the excitation energy (circles). 
(e) SFRS intensity resonance profile for the electron spin-flip at $B_{\rm V}=7$~T. All data are measured at $T=1.6$~K. 
}
\label{fig:2}
\end{center}
\end{figure*}

The schematics of the applied experimental geometries are shown in Figure~\ref{fig:2}a. In the Faraday geometry the  magnetic field is parallel to the light $k$-vector ($\mathbf{B}_\text{F} \parallel \mathbf{k}$), which in turn  is parallel to the crystal $c$-axis ($\mathbf{k} \parallel  c$). In the studied (PEA)$_2$PbI$_4$, SFRS signal is absent in the pure Faraday geometry, because the carrier spin states $+1/2$ and $-1/2$ are not mixed by magnetic field and, thus, the spin-flip process is suppressed. Such mixing already occurs for small tilt angles, e.g. at $\theta = 10^\circ$, and then indeed SFRS signals become  pronounced. We refer to this geometry as "close-to-Faraday geometry", where the Zeeman splitting is dominated by the $g$-factor component along the $c$-axis ($g_{\rm e(h),c}$). The SFRS spectrum measured at $B=9.4$~T with the excitation laser tuned to the exciton resonance at $2.345$~eV, is shown in Figure~\ref{fig:2}b. The spin-flip lines are more pronounced in the anti-Stokes spectral range, as there the contribution of background photoluminescence is minimized.    

Four spin-flip lines in the SFRS spectrum labeled by $E_\text{h}$, $E_\text{e+h}$, $E_\text{e}$, and $E_\text{e-h}$ are seen in Figure~\ref{fig:2}b. They are absent at zero magnetic field, as expected from the vanishing Zeeman splitting so that the spin-flip lines coincide with the exciting laser energy. With increasing magnetic field the lines shift linearly from the laser energy (referred to as zero). The details of their shifts are shown in Figures~\ref{fig:$g$-factor}a,b,c. The $E_\text{h}$ line is only detected in high magnetic fields (see the black circles in Figure~\ref{fig:$g$-factor}c) due to its small shift amounting to only $-0.078$~meV at 9.4~T, which is associated with the spin-flip of the hole having $|g_\text{h,c}|=0.13$. The electron spin-flip line shows a much larger shift of $E_\text{e}=-1.137$~meV at $B=9.4$~T. Its magnetic field dependence in Figure~\ref{fig:$g$-factor}a allows us to evaluate $|g_\text{e,c}| = 2.11$, agreeing well with $g_\text{e,c}=+2.05\pm0.05$ determined from time-resolved Kerr rotation on the same sample~\cite{kirstein2022c}. 

Identification of the hole and electron SFRS lines is done by comparing their $g$-factors with the universal dependence of the carrier $g$-factors on the band gap energy, that has recently been established for bulk lead halide perovskites~\cite{kirstein2022}. According to this dependence for materials with band gap energies around 2.3~eV, $g$-factor values of $g_\text{e}  \approx  +2.0$ and $g_\text{h} \approx +0.7$ are predicted. We expect some, but not drastic deviations from this dependence for the 2D perovskites. Therefore, we assign the spin-flip line with the larger shift to the resident electron with positive sign of the $g$-factor.  

Another approach to distinguish electrons from holes is based on their interactions with the nuclear spin system, since in lead halide perovskites the hole-nuclei interaction is about five times stronger than the electron-nuclei one~\cite{kirstein2021}. This also leads to a much stronger dynamic nuclear polarization (DNP) by the holes. In Section~\ref{sec:DNP} we show that DNP can be detected with the SFRS technique and that its effect is considerable for the $E_\text{h}$ line. Note that it is absent for the $E_\text{e}$ line, in accordance with our assignment. From this experiment, the sign of the hole $g$-factor  can be unambiguously determined. For the studied (PEA)$_2$PbI$_4$, the hole $g$-factor is negative, i.e. $g_\text{h,c}=-0.13$.

The electron line $E_\text{e}$ has two satellites, $E_\text{e-h}$ and $E_\text{e+h}$, which are shifted by the hole Zeeman splitting. The slopes in their magnetic field dependence give $|g_\text{e-h,c}|=2.25$ and $|g_\text{e+h,c}|=1.94$, see Figure~\ref{fig:$g$-factor}b. These lines are provided by double spin-flip processes, in which simultaneously electron and hole spin-flips are involved. This is a rather unusual SFRS process. Double electron spin-flip was found experimentally in 1972 for an exciton interacting with two donor-bound electrons in CdS~\cite{Scott1972}, later in ZnTe~\cite{Oka1981}, and recently for two localized electrons interacting with the same exciton in CdSe colloidal nanoplatelets~\cite{Kudlacik2020}. The according theoretical consideration can be found in Refs.~\citenum{Economou1972,Rodina2020_NPL,Rodina2022}. The $E_\text{e-h}$ and $E_\text{e+h}$ line shifts are much larger compared to the $E_\text{h}$ line, and can therefore be resolved in a larger range of magnetic fields starting from 4~T. We use the difference between the double spin-flip line $E_\text{e+h}$ and the electron line $E_\text{e}$ to evaluate the hole Zeeman splitting vs magnetic field in Figure~\ref{fig:$g$-factor}c. Note that the full width at half maximum, taken from the Gaussian fit of the electron spin-flip line, is about 7 times larger ($50~\mu$eV) than that for the hole ($7~\mu$eV), which indicates a broader electron $g$-factor dispersion.

Commonly, SFRS signals have a pronounced polarization dependence, caused by the optical selection rules and the involved scattering mechanisms. The spectrum shown in Figure~\ref{fig:2}b is measured in a cross circularly-polarized configuration with $\sigma^+$ polarized excitation and $\sigma^-$ polarized detection. The other polarization configurations are shown in the Supporting Information, Figure~S1a, for both the anti-Stokes and Stokes spectral ranges. Surprisingly, the polarization dependence is weak. Possible reasons for that are discussed in Section~\ref{sec:mechanism}.

\subsection{Spin-flip Raman scattering in Voigt geometry} 
\label{sec:Voigt}

In order to determine the in-plane components of the electron and hole $g$-factors, we perform SFRS measurements in the Voigt geometry, where $\mathbf{B}_\text{V} \perp \textbf{k}$, $\mathbf{B}_\text{V} \parallel (a,b)$ and $\theta=90^\circ$. In Figure~\ref{fig:2}c the SFRS spectrum for linearly co-polarized excitation and detection in the Voigt geometry at $B_\text{V}=10~$T is shown. This geometry is favorable for SFRS experiments because the spin states are mixed by the perpendicular magnetic field, facilitating an efficient spin-flip process. The comparison of the SFRS spectra in Figures \ref{fig:2}b and \ref{fig:2}c shows that the SFRS intensity in the Voigt geometry is about five times higher than in the Faraday geometry. In the Voigt geometry the Raman shift of the electron $E_\text{e}$ line corresponds to $|g_\text{e,(a,b)}|=2.50$, see also Figure~\ref{fig:$g$-factor}d. This value is in good agreement with $g_\text{e,(a,b)}=+2.45\pm0.05$ measured by time-resolved Kerr rotation~\cite{kirstein2022c}. The slope of the linear fit to the double spin-flip $E_\text{e+h}$ Raman shift corresponds to  $|g_\text{e+h,(a,b)}|$~$=$~$1.98$, see Figure~\ref{fig:$g$-factor}e. As the $E_\text{h}$ line cannot be resolved in this geometry, we calculate its Raman shift from the shift difference between $E_\text{e+h}$ and $E_\text{e}$. From the data in different magnetic fields we determine the hole $g$-factor $g_\text{h,(a,b)}=-0.51$ (Figure~\ref{fig:$g$-factor}f). Note that the $E_\text{e-h}$ line cannot be well detected in Voigt geometry. The SFRS spectra are measured in different configurations of linear polarization, however, a noticeable influence of selection rules is not found, more information is provided in the Supporting Information, Figure~S1b.      

It is worth pointing out that the amplitude of the spin-flip lines is sensitive to temperature. We show in the Supporting Information, Figure~S2, that the electron SFRS line amplitude decreases for temperatures exceeding 5~K and becomes weak above 16~K. We suggest that thermal delocalization of the resident electrons is the mechanism that reduces the efficiency of the SFRS process. The estimated  activation energy is about 2.1~meV.

\subsection{Resonance profile of spin-flip Raman scattering}
\label{sec:Spectral}

The SFRS intensity has a strong spectral dependence on the laser photon energy as shown in Figure~\ref{fig:2}e for the $E_\text{e}$ line, measured in the Voigt geometry at $B_{\rm V}=7$~T. The maximum of the resonance profile coincides with the free exciton energy in the reflectivity spectrum shown in Figure~\ref{fig:2}d, where also the PL line is given for comparison. This highlights the key role of the exciton in the SFRS process. The exciton resonantly enhances the laser excitation and scattering through the interaction with resident carriers, whose spin-flip Raman shift is measured. Similar results were reported for CdTe/(Cd,Mg)Te quantum wells with a low density of resident electrons~\cite{Debus2013_CdTeQW} and for singly-charged (In,Ga)As/GaAs quantum dots~\cite{Debus2014_SFRS_QDs}. In Figure~\ref{fig:2}d, the electron $g$-factor is shown as function of the excitation energy. It remains constant across the investigated energy range.

\subsection{$g$-factors of electrons and holes and their anisotropy}
\label{sec:factors}

\begin{figure*}[hbt]
\begin{center}
\includegraphics[width=17.8cm]{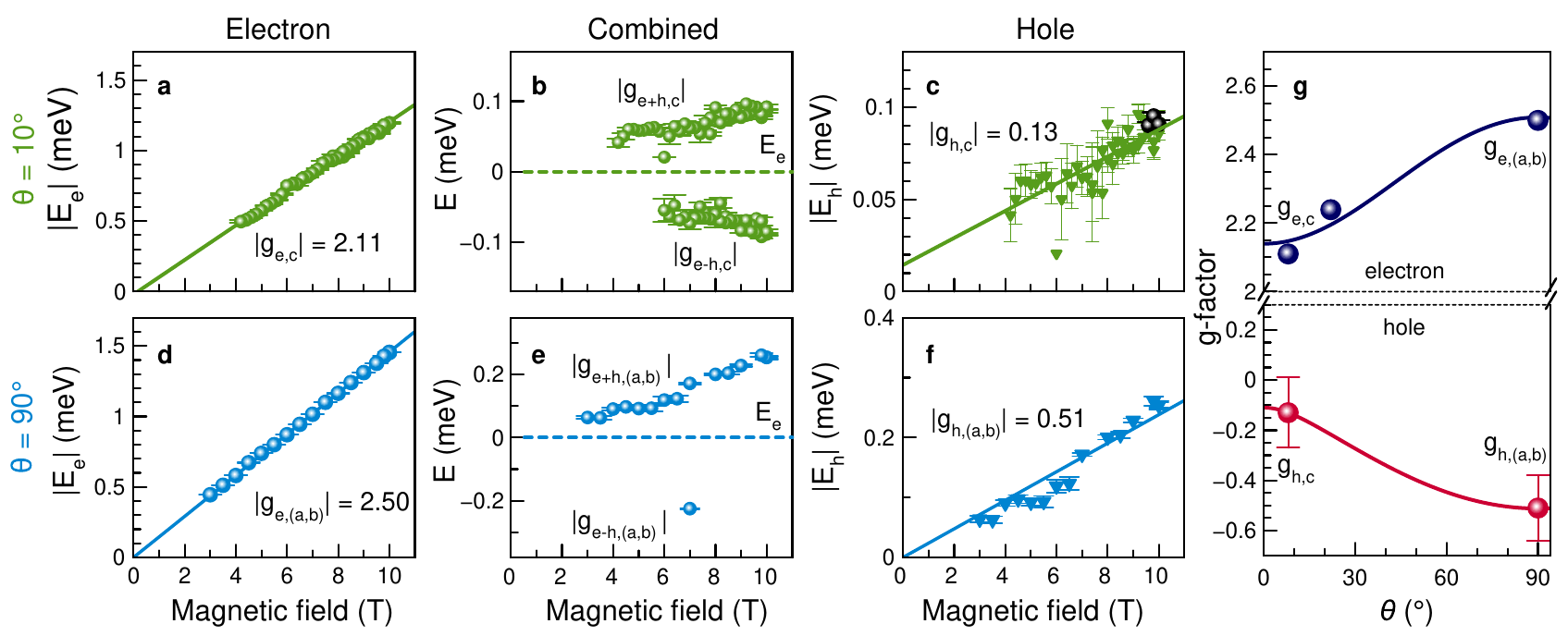}
\caption{Raman shifts of the spin-flip lines in magnetic field and evaluated carrier $g$-factors in (PEA)$_2$PbI$_4$.
Magnetic field dependences of the anti-Stokes SFRS shifts measured in the close-to-Faraday ($\theta=10^\circ$) and Voigt ($\theta=90^\circ$) geometries. 
(a, d) Electron $E_\text{e}$ shift. 
(b, e) Double spin-flip $E_\text{e+h}$ and $E_\text{e-h}$ shifts after subtracting the shift of $E_\text{e}$.
(c, f) Hole $E_\text{h}$ shift evaluated from the double spin-flip shifts (colored triangles). Black dots in panel (c) give direct measurements of the hole spin-flip shift of the $E_\text{h}$ line.  In all panels, $B$-linear fits are shown by the solid lines. 
(g) $g$-factor anisotropy for the tilt angle $\theta$ tuned between Faraday and Voigt geometry. Lines are fits with Equation~\eqref{eqn:$g$-factor anistropy}. All measurements presented here are performed at $T=1.6$~K with $E_{\text{exc}}=2.345$~eV laser photon energy using $P=7.5$~W/cm$^2$ excitation power.
\label{fig:$g$-factor}}
\end{center}
\end{figure*}

The electron and hole $g$-factors can be precisely determined from the Raman shifts of the respective lines at different magnetic fields. The results in the close-to-Faraday and Voigt geometries are presented in Figure~\ref{fig:$g$-factor}. The electron $g$-factors are taken from the shift of the $E_\text{e}$ line, see Figures~\ref{fig:$g$-factor}a,d. Figures~\ref{fig:$g$-factor}b,e illustrate the shifts of the double spin-flip lines  $E_\text{e+h}$ and  $E_\text{e-h}$, which are plotted relative to the $E_\text{e}$ line shift. The differences between the $E_\text{e}$ and the double spin-flip line shifts correspond to the hole Zeeman splitting, shown in Figures~\ref{fig:$g$-factor}c,f. Only in high magnetic fields the $E_\text{h}$ shift can be directly measured. The corresponding values of the $g$-factors are given in the panels of Figure~\ref{fig:$g$-factor} and are also collected in Table~\ref{Tab:1}, in which the signs of the $g$-factors are given.

\begin{table*}[hbt]
\begin{center}
\begin{tabular}{|c|p{1,1cm}|p{1,1cm}|p{1,1cm}|p{1,1cm}||p{1,1cm}|p{1,1cm}|p{1,1cm}|p{1,1cm}|}
\hline
      & \multicolumn{4}{c||}{close to Faraday geometry, $\theta = 10^\circ$}       & \multicolumn{4}{c|}{Voigt geometry, $\theta = 90^\circ$} \\ \hline
      & $g_\text{e,c}$ & $g_\text{h,c}$& $g_\text{e-h,c}$ & $g_\text{e+h,c}$ & $g_\text{e,(a,b)}$ & $g_\text{h,(a,b)}$ & $g_\text{e-h,(a,b)}$ & $g_\text{e+h,(a,b)}$\\ \hline
       SFRS & +2.11  &  $-0.13$ & +2.25 & +1.94 & +2.50 & $-0.51$ & +2.93 & +1.98 \\ \hline
       TRKR~\cite{kirstein2022c} & +2.05     &  &  &  & +2.45 &  &  & \\ \hline
\end{tabular}
\caption{Overview of the $g$-factors in the close-to-Faraday geometry ($\theta=10^\circ$) and in the Voigt geometry ($\theta=90^\circ$) for (PEA)$_2$PbI$_4$ measured by SFRS and time-resolved Kerr rotation~\cite{kirstein2022c}. The measurement accuracy is $\pm0.05$ in all cases. The exciton Zeeman splitting measured by magneto-reflectivity in Faraday geometry is $g_{\text{X,c}}=+1.6\pm0.1$. 
\label{Tab:1}}
\end{center}
\end{table*}

The anisotropy of the carrier $g$-factors is inherent for 2D structures and originates from the reduced symmetry of the band structure. For the studied (PEA)$_2$PbI$_4$ sample the anisotropy is shown in Figure~\ref{fig:$g$-factor}g, where the experimental data for the close-to-Faraday and Voigt geometries are complemented by measurements at the magnetic field tilt angle of $\theta=22^\circ$. The angular dependence of the $g$-factor can be described by   
\begin{equation}
g(\theta)=\sqrt{(g_\text{c}\cos \theta)^2 + (g_\text{(a,b)}\sin \theta)^2} \,.
\label{eqn:$g$-factor anistropy} 
\end{equation}
The electron $g$-factor anisotropy measured by TRKR in a vector magnet with smaller steps of the tilt angle can be found in Ref.~\cite{kirstein2022c}. 
It is interesting to note that the anisotropies of $g_\text{e}$ and $g_\text{h}$ almost compensate each other, so that their sum stays nearly isotropic,  $g_\text{e,c}+g_\text{h,c}$~$=$~$+1.98$ and $g_\text{e,(a,b)}+g_\text{h,(a,b)}$~$=$~$+1.99$. A similar behavior was recently found for bulk CsPbBr$_3$ crystals~\cite{kirstein2022}. 

The $g$-factor of the bright exciton in lead halide perovskites is the sum of the carrier $g$-factors:
\begin{equation}
g_{\rm X}=g_{\rm e}+g_{\rm h} \,.
\label{eqn:gX} 
\end{equation}
Therefore, the exciton $g$-factor in 2D (PEA)$_2$PbI$_4$ should be nearly isotropic despite a clear crystal anisotropy. In fact, $g_{\rm X}$ may deviate from the relation \eqref{eqn:gX}, as some $g$-factor renormalization can occur at finite carrier $k$-vectors in the exciton. It is instructive to check this relation for (PEA)$_2$PbI$_4$. Here $g_\text{e,c}+g_\text{h,c}=+1.98$ can be compared with the $g_{\rm X,c}=+1.6$ measured by magneto-reflectivity (Figure~\ref{fig:1}c). Indeed, the exciton $g$-factor is about 0.4 smaller than the sum. We attribute this difference to the large exciton binding energies in the 2D (PEA)$_2$PbI$_4$ perovskite. Further model calculations are needed to identify the involved mechanisms.

\subsection{Dynamic nuclear polarization }
\label{sec:DNP}

The spin dynamics of electrons and holes in semiconductors are strongly influenced by their hyperfine interaction with the nuclear spin system~\cite{OpticalOrientation}. In conventional III-V and II-VI semiconductors the conduction band electrons with s-type wave functions have stronger interaction with the nuclei spins compared to the valence band holes with p-type wave functions. This situation is reversed in lead halide perovskites, where the Pb ions greatly contribute to the states around the band gap~\cite{kirstein2021,kirstein2022b}. Their p-orbitals form the conduction band, while the s-orbitals contribute to the valence band. As a result, the hyperfine interaction of the holes is about 5 times stronger than that of electrons. Similar properties are expected for (PEA)$_2$PbI$_4$.

\begin{figure*}[hbt]
\begin{center}
\includegraphics[width=17.8cm]{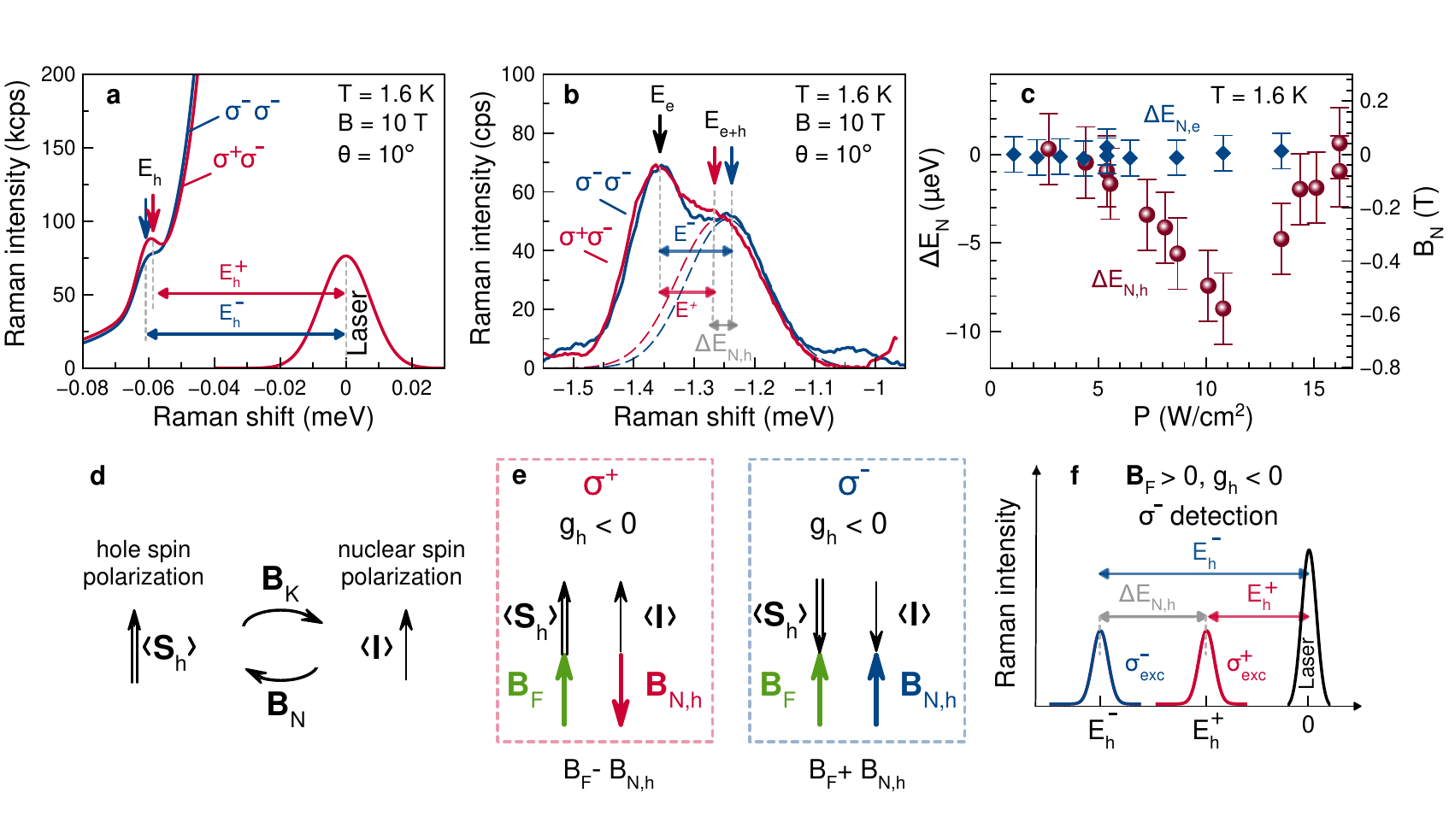}
\caption{ Dynamic nuclear polarization of holes detected via SFRS in (PEA)$_2$PbI$_4$.
(a) Raman shift of the hole $E_\text{h}$ line measured in $\sigma^-\sigma^-$ and $\sigma^+\sigma^-$ polarization for $P=5.1~$W/cm$^2$ excitation power.
(b) Raman shift of the electron $E_\text{e}$ and double spin-flip $E_\text{e+h}$ lines measured in different polarizations for $P=13.5~$W/cm$^2$.
(c) Power density dependences of the energy splitting $\Delta E_\text{N}=E^{+}-E^{-}$ for the electron and hole shifts from panel (b). Right axis gives the corresponding Overhauser field $B_\text{N}$. 
(d) Schematic illustration of the hyperfine interaction in a carrier-nuclei spin system. The effective Knight field ($B_\text{K}$) of the spin polarized holes $\langle\mathbf{S}_\text{h}\rangle$ acts on the nuclear spin system. The average nuclear spin polarization $ \langle \mathbf{I} \rangle$ acts back via the Overhauser field ($B_\text{N,h}$) on the hole spin.
(e) $\mathbf{B}_\text{N,h}$ orientation scheme for $g_\text{h}<0$ with $\sigma^{+}$ or $\sigma^{-}$ excitation in an external magnetic field applied in the Faraday geometry.  
(f) Schematic Raman spectrum highlighting the effect of DNP on the Raman shift. The difference between the Raman line shifts for $\sigma^+$ and $\sigma^-$ excitation is proportional to $2B_\text{N,h}$.
\label{fig:Overhauser}}
\end{center}
\end{figure*}

The hyperfine interaction of carriers with the nuclei can be assessed by the effect of dynamic nuclear polarization (DNP). Spin polarized carriers, which can be generated by circularly polarized light using optical orientation, generate the Knight field ($\mathbf{B}_{\rm K}$) that acts as an effective magnetic field on the nuclear spins. Thereby the carrier spin polarization can be transferred to the nuclear spin system, so that it becomes polarized. In turn, the polarized nuclear system induces the Overhauser field ($\mathbf{B}_{\rm N}$) that acts on the carrier spins and changes their Zeeman splitting, see scheme in Figure~\ref{fig:Overhauser}d. Details of DNP model description are given in Ref.~\cite{kirstein2021}. Here for simplicity, we will consider only the hole contribution, as we found experimentally that in (PEA)$_2$PbI$_4$ the holes are dominant in polarizing the nuclear spins.  

The spin polarization of optically oriented holes $\langle \mathbf{S}_\text{h} \rangle$ is transferred to the nuclei and induces the average nuclear spin polarization $\langle \mathbf{I} \rangle$ given by
\begin{equation}
     \langle \mathbf{I} \rangle = l \frac{4I(I+1)}{3} \frac{\mathbf{B} (\mathbf{B} \langle \mathbf{S}_\text{h} \rangle)}{B^2} \, .
\end{equation} 
Here $l$ is the leakage factor characterizing DNP losses due to relaxation processes and $I$ is the nuclear spin. The nuclear spin polarization builds up along the direction of projection of $\langle\mathbf{S}_\text{h}\rangle$ onto the magnetic field, and therefore its direction can be controlled by the light helicity, which determines the $\langle\mathbf{S}_\text{h}\rangle$ orientation. The nuclear spin polarization can be converted into the Overhauser field:
\begin{equation}
\label{eq:BN}
    \mathbf{B}_{\text{N,h}}=\frac{\alpha A_\text{h} \langle \mathbf{I} \rangle}{g_\text{h}\mu_{\rm B}}
\end{equation}
with the positive hyperfine coupling constant for holes $A_\text{h} > 0$. Here $\alpha$ is abundance of nuclear isotops with nonzero spin. The sign of $\mathbf{B}_\text{N,h}$ is determined by the sign of the hole $g$-factor, which offers an experimental tool for evaluating the $g_\text{h}$ sign. In Figure~\ref{fig:Overhauser}e, diagrams for the orientation possibilities of the magnetic and effective fields as well as the spin polarization are given for $\sigma^+$ and $\sigma^-$ circular polarized excitation. Here, we take $g_\text{h}<0$. Depending on the $g_\text{h}$ sign, $\mathbf{B}_\text{N,h}$ can increase or reduce the hole Zeeman splitting induced by an external magnetic field. Therefore, it changes the spin-flip Raman shift and can be detected experimentally by SFRS. For example, we have demonstrated that for (In,Ga)As/GaAs quantum dots~\cite{Debus2015_SFRS_nuclei}.

In order to examine the DNP in (PEA)$_2$PbI$_4$, we apply the SFRS technique in close-to-Faraday geometry ($\theta=10^\circ$). The signal is detected in $\sigma^-$ circular polarization, while the excitation polarization was set to either $\sigma^+$ or $\sigma^-$. As shown in Figure~\ref{fig:Overhauser}a the shift of the hole spin-flip line is larger in the $\sigma^-\sigma^-$ configuration (we label the shift as $E^-_\text{h}$ shown by the blue arrow) than in the $\sigma^+\sigma^-$ configuration ($E^+_\text{h}$, the red arrow). Their difference scales with twice the Overhauser field which thus can be extracted from the relation 
\begin{equation}
\label{eq:DN}
    \Delta E_{\text{N,h}}= |E^{+}_{\text{h}}| - |E^{-}_{\text{h}}|=2 |g_\text{h,c}| \mu_B B_\text{N,h}\, ,
\end{equation}
as sketched in Figure~\ref{fig:Overhauser}f. The  difference $\Delta E_\text{N,h} =|E_\text{h}^+|-|E_\text{h}^-|=-1.6~\mu$eV measured at $P=5.1~$W/cm$^2$ corresponds to $B_\text{N,h}=-0.11$\,T calculated using $|g_\text{h,c}|=0.13$. The negative sign of $B_\text{N,h}$ means that $g_\text{h,c}<0$, see Equation~\eqref{eq:BN}.

The nuclear-induced shift can be seen even more clearly in the double spin-flip line $E_{\rm e+h}$, as for it the background contribution of the scattered laser light is strongly reduced, see Figure~\ref{fig:Overhauser}b. Note that the DNP shift is absent for the electron Zeeman splitting, as the $E_{\rm e}$ shift is the same for the $\sigma^-\sigma^-$ and $\sigma^+\sigma^-$ configurations. However, the shift of the $E_{\rm e+h}$ line from $E_{\rm e}$ varies. For the used excitation density of 13.5~W/cm$^2$ the energy splitting of these lines amounts to $\Delta  E_\text{N,h}=-5.5~\mu$eV and, therefore, to $B_\text{N,h}=-0.36$~T.    

The excitation density dependences of $\Delta  E_\text{N}(P)$ and $B_\text{N}(P)$ for electrons and holes are shown in Figure~\ref{fig:Overhauser}c.
As already noted, any effect on the electrons is absent, reflecting the expected weak hyperfine interaction in the 2D (PEA)$_2$PbI$_4$ perovskite. The energy splitting is pronounced for holes, for which the $\Delta  E_\text{N,h}$ value increases up to $-8.7~\mu$eV, which corresponds to $B_\text{N,h}=-0.6~$T at $P=10.5~$W/cm$^2$. For a further excitation density increase the energy splitting becomes weaker, which we assign to heating of the nuclear spin system. 

The results of this section show that the SFRS technique is a valuable tool for providing insight into the central spin problem of a carrier spin placed in a nuclear spin bath in 2D perovskite materials and also for studying perovskite bulk crystals and nanocrystals. 
 
\subsection{Discussion} 
\label{sec:mechanism}

Let us discuss the mechanisms that can be responsible for the observed spin-flip processes. Recently a detailed theoretical analysis of the spin-flip Raman scattering processes involving excitons and resident charge carriers in perovskite semiconductors was published~\cite{Rodina2022}. The model considerations were made for bulk perovskites with cubic symmetry. Three mechanisms for the observation of carrier spin-flips were suggested that can explain both the single and double spin-flip processes. The first mechanism is the resonant excitation of a localized exciton, followed by its exchange interaction with a resident electron and/or hole. The second mechanism involves a biexciton as an intermediate state of the SFRS. In this case, spin-flip shifts by the energies of $E_\text{e+h}$ and $E_\text{e+h}/2$ are expected, but the process with $E_\text{e-h}$ is forbidden. We observe, however, in the experiment the $E_\text{e-h}$ process, which allows us to exclude the biexciton scenario. The third mechanism is the direct excitation of propagating exciton-polaritons, their scattering on resident carriers, and the conversion of the polaritons into secondary photons at the sample boundary. In 2D perovskites the semiconductor layers are electronically decoupled from each other which prevents exciton-polariton motion along the $c$-axis. Therefore, we suggest that the exciton-polariton mechanism can be neglected for them. 

Our experimental results on the spectral dependence of the SFRS intensity clearly show the exciton involvement. In turn, the Raman shift values and their anisotropy allow us to refer them to resident carriers, which interact with the exciton acting as a mediator of the SFRS. The considerations of Ref.~\citenum{Rodina2022} predict pronounced polarization dependencies for the carrier spin-flip lines. So far, the reason why such dependencies are not observed here is unclear, see the Supporting Information, Figure~S1. Future analysis accounting for the reduced symmetry and mixing of the bright exciton states in 2D perovskites might clarify this property. Note that the 2D perovskites are similar 2D CdSe colloidal nanoplatelets, but for the nanoplatelets we do observe pronounced polarization dependences of the spin-flip lines~\cite{Kudlacik2020}. The SFRS theory for the CdSe nanoplatelets is presented in Ref.~\citenum{Rodina2020_NPL}. Here, the violation of polarization selection rules caused by the in-plane anisotropy of the nanoplatelets, which induces exciton mixing and splitting, and by the finite Zeeman splitting of the intermediate state were analyzed. 
 
We have measured the exciton $g$-factor in (PEA)$_2$PbI$_4$ of $g_\text{X,c} = +1.6$ from the exciton Zeeman splitting in magneto-reflectivity, see Figure~\ref{fig:1}c. It is in agreement with reported experimental data for 2D perovskite excitons measured in pulsed magnetic fields up to 60~T, where either magneto-transmission or magneto-reflectivity were used. The exciton $g$-factor of about 1.2 was measured for (PEA)$_2$PbI$_4$~\cite{dyksik2020}. For the similar 2D perovskite (C$_6$H$_{13}$NH$_3$)$_2$PbI$_4$ it amounts to about 1.5~\cite{Baranowski2019} and $+1.80$~\cite{kataoka1993}. For (C$_{10}$H$_{21}$NH$_3$)$_2$PbI$_4$ it is $+1.42$~\cite{hirasawa1993}, and for the halogen substitution for Br in (C$_4$H$_{9}$NH$_3$)$_2$PbBr$_4$ the $g$-factor is $+1.2$~\cite{tanaka2005}.   

Another type of spin-flip Raman scattering process was observed in the (C$_4$H$_9$NH$_3$)$_2$PbBr$_4$ 2D perovskite~\cite{Ema2006}. These experiments were performed at zero magnetic field, where the Raman shift by the large value of $28-32$~meV arises from the scattering of a bright exciton with parallel electron and hole spins into a dark exciton state with antiparallel spins. This shift corresponds to the bright-dark exchange splitting of the excitons and not to the Zeeman splitting of their states.

\section{Conclusions}

We have investigated the spin properties of two-dimensional (PEA)$_2$PbI$_4$ perovskites using spin-flip Raman scattering. We have found spin-flip signals from resident electrons and holes, as well as their combinations. This has allowed us to measure the Land\'e factors and their anisotropy. The anisotropy of the electron and hole $g$-factors is complementary, so that their sum corresponding to the bright exciton $g$-factor remains isotropic. Also, hyperfine hole-nuclei interaction is demonstrated in 2D perovskites by means of the dynamic nuclear polarization. Due to the small $g$-factor of the hole, we were able to achieve an Overhauser field value of $B_\text{N,h} = 0.6$\,T. The direction of $\mathbf{B}_\text{N,h}$ allows us to unambiguously determine the negative sign of the hole $g$-factor. We are convinced that similar effects as those observed can manifest themselves in the large class of 2D lead halide perovskites with different numbers of layers, as well as in perovskites with various organic cations.

\section{Experimental Section}

\subsection{Samples}
We studied Ruddlesden-Popper type two-dimensional  (PEA)$_2$PbI$_4$ perovskites that consist of a stack of monolayers formed by corner-shared PbI$_6$ octahedra. The monolayers are separated by van der Waals-bonded pairs of PEA [phenethylammonium (C$_6$H$_5$)C$_2$H$_4$NH$_3$] molecules. Details of the synthesis are given in the Supporting Information, S1, and in Ref.~\citenum{kirstein2022c}. Due to the quantum confinement of electrons and holes the band gap energy of (PEA)$_2$PbI$_4$ is $2.608$~eV at $T=2$~K~\cite{dyksik2020}. This value considerably exceeds the band gap energy of $A$PbI$_3$ lead iodine archetype bulk crystals with $E_g =1.5-1.7$~eV, where $A=$ Cs$^+$, MA$^+$, or FA$^+$. The reduced dimensionality and the dielectric enhancement effect, provided by the contrast in dielectric constants between the perovskite monolayers and the PEA, strongly increase the exciton binding energy in (PEA)$_2$PbI$_4$ to $260$~meV~\cite{dyksik2020,Baranowski2022} in comparison to $16$~meV in bulk MAPbI$_3$ perovskite~\cite{Galkowski2016}.

\subsection{Experimental details}

\subsubsection{Spin-flip Raman scattering (SFRS) spectroscopy} 
The SFRS technique enables one to directly measure the Zeeman splitting of the electron and hole spins from the spectral shift of the scattered light from the laser photon energy.  Resonant excitation of the exciton strongly increases the SFRS signals, allowing one to measure resident electrons and holes interacting with an exciton. For optical excitation we used a tunable single-frequency continuous wave Ti:Sapphire laser (Matisse DS) equipped with a MixTrain module from SIRAH. The emitted photon energy was tuned around 530~nm (spectral range $2.33-2.36$~eV), provided by the sum frequency of the Ti:Sapphire laser operating around $720$~nm and a fiber laser emitting at $1950$~nm. The actual wavelength was measured and monitored by a fiber-coupled high-resolution wavemeter (HighFinesse WSU). The laser power after the MixTrain module was generally set to $0.7$~mW. The diameter of the laser spot on the sample was 180~$\mu$m in diameter resulting in an excitation density of $P=2.75$~W/cm$^2$, if not specified otherwise. The linear or circular polarizations of the laser beam and the Raman signal were set and analyzed by combinations of $\lambda/2$ or $\lambda/4$ wave plates and a Glan-Thompson prism, positioned in the excitation and detection paths. The linear polarizations are denoted as V (vertical) and H (horizontal), and the circular polarizations as $\sigma^{+}$ and $\sigma^{-}$.

The Raman signals were measured in backscattering geometry. The light scattered from the sample was dispersed by a $1$~m double monochromator (Yobin-Yvon U1000) equipped with a Peltier-cooled GaAs photomultiplier providing a spectral resolution of $0.8$~$\mu$eV. This allows us to measure $g$-factors with an accuracy of $0.05$. To protect the photomultiplier from the highly intense laser light, a neutral density filter is placed in the detection path while recording the laser. The SFRS measurements were performed at the low temperature of $T=1.6$~K with the sample immersed in pumped liquid helium. Magnetic fields up to $10$~T generated by a superconducting split-coil solenoid were applied. In Figure~\ref{fig:2}a the used experimental geometries are shown. The sample axes $a$ and $b$ are in-plane, while the $c$-axis is out-of-plane. In the Faraday geometry, the magnetic field is parallel to the light wave vector $k$ and to the $c$-axis ($\mathbf{B}_\text{F} \parallel \textbf{k}$, $\mathbf{B}_\text{F} \parallel c$ and $\theta=0^\circ$). In the Voigt geometry the field is perpendicular to these vectors ($\mathbf{B}_\text{V} \perp \textbf{k}$, $\mathbf{B}_\text{V} \parallel (a,b)$ and $\theta=90^\circ$). In our experiments we did not distinguish between the orientations of the $a$- and $b$-axes, and therefore the measured values are averaged over their random orientations, while the differences between the two axes are not expected to be large. The angle $\theta$ between the $c$-axis and the magnetic field specifies the tilt of the field directions as shown in the bottom diagram, where $\textbf{k} \parallel \mathbf{B}$ is kept.
\\ 

\subsubsection{Photoluminescence (PL)}
Nonresonant continuous wave excitation with the photon energy of 2.412~eV and an excitation density of $P=10.8$~W/cm$^2$ was used for the PL measurements performed at $T=1.6$~K. The PL was detected by the same $1$~m double monochromator (U1000) and GaAs photomultiplier, as in the SFRS measurements.    

\subsubsection{Time-resolved photoluminescence (TRPL)}
For the TRPL measurements, a pulsed excitation laser was used (pulse duration of $10$~ns, pulse repetition rate of $800$~Hz, photon energy of $3.493$~eV, and excitation density of $P=3$~W/cm$^2$). The PL was detected again with the Peltier-cooled GaAs photomultiplier coupled to the U1000. The time resolution of the recombination dynamics was provided by a time-of-flight electronic board (Fast ComTec MCS6A), which has a nominal time resolution of 100~ps. In our experiment the time resolution of 10~ns was limited by the laser pulse duration.

\medskip
\textbf{Supporting Information} \par 
 Correspondence and requests for material should be addressed to D.R.Y. (dmitri.yakovlev@tu-dortmund.de).

\medskip
\textbf{Acknowledgements} \par 
We are thankful to A. V. Rodina, E. L. Ivchenko, M. O. Nestoklon, M. M. Glazov, and I. V. Kalitukha for fruitful discussions. We acknowledge the financial support by the Deutsche Forschungsgemeinschaft in the frame of the Priority Programme SPP 2196 (Project YA 65/26-1) and the International Collaboration Research Center TRR160 (Projects A1 and B2). The work of O.H. and M.V.K. was financially supported by the Swiss National Science Foundation (grant agreement 186406, funded in conjunction with SPP2196 through DFG-SNSF bilateral program) and by the ETH Z\"urich through the ETH+ Project SynMatLab: Laboratory for Multiscale Materials Synthesis.

\subsection*{Competing interests}
The authors declare no competing interests.

\medskip

\textbf{ORCID} \\
Carolin Harkort:      0000-0003-1975-9773 \\
Dennis Kudlacik:      0000-0001-5473-8383 \\
Nataliia E. Kopteva:  0000-0003-0865-0393 \\
Dmitri R. Yakovlev:   0000-0001-7349-2745 \\
Marek Karzel:         0000-0002-1939-5191\\
Erik Kirstein:        0000-0002-2549-2115 \\
Oleh Hordiichuk:      0000-0001-7679-4423 \\
Maksym~V.~Kovalenko:  0000-0002-6396-8938 \\
Manfred Bayer:        0000-0002-0893-5949\\

\bibliographystyle{MSP} 

\newpage
\clearpage

\onecolumngrid

\begin{center}
	\textbf{\Huge Supporting Information:}




\vspace{3mm}	
	\textbf{\Large Spin-flip Raman scattering on electrons and holes in two-dimensional (PEA)$_2$PbI$_4$ perovskites}
 
\end{center}	





\twocolumngrid

\setcounter{equation}{0}
\setcounter{figure}{0}
\setcounter{table}{0}
\setcounter{section}{0}
\setcounter{page}{1}
\renewcommand{\thepage}{S\arabic{page}}
\renewcommand{\theequation}{S\arabic{equation}}
\renewcommand{\theHequation}{S\arabic{equation}}
\renewcommand{\thefigure}{S\arabic{figure}}
\renewcommand{\theHfigure}{S\arabic{figure}}
\renewcommand{\thetable}{S\arabic{table}}
\renewcommand{\theHtable}{S\arabic{table}}
\renewcommand{\thesection}{S\arabic{section}}
\renewcommand{\theHsection}{S.\arabic{section}}
\renewcommand{\thefootnote}{\fnsymbol{footnote}} 

\renewcommand{\theHfigure}{\Alph{section}.\arabic{figure}}
\renewcommand{\theHequation}{\Alph{section}.\arabic{equation}}
\renewcommand{\theHfootnote}{S.\Alph{section}.\Alph{footnote}} 


\section{Preparation of the (PEA)$_2$PbI$_4$ thin films} 

The monolayer-thick 2D hybrid organic-inorganic lead halide perovskites have the stoichiometry $A_m$Pb$X_4$, with the halide $X$ of the corner shared [PbX$_6$]$^{4-}$ octahedra forming the monolayers, the bigger organic molecule $A$ separating these monolayers, and $m=1$ for the divalent Dion-Jacobson or $m=2$ for the monovalent Ruddlesden-Popper configuration~\cite{SI_mao2018}. While in the bulk case $A$ is limited by the tolerance factor to \{Cs, formamidinium, methylammonium\}, in the case of 2D perovskites $A$ can be chosen from a plethora of organic molecules~\cite{SI_mao2018}. 

For this study Ruddlesden-Popper $m=2$ thin films of PEA [phenethylammonium (C$_6$H$_5$)CH$_2$CH$_2$NH$_3$] lead iodide, (PEA)$_2$PbI$_4$, were chosen. For the synthesis, (PEA)I and PbI$_2$ were dissolved in 
N,N-dimethylformamide in 2:1 molar ratio to obtain a solution of 8.3\,wt.\% concentration of (PEA)$_2$PbI$_4$. The solution was spin-coated on glass substrates (the substrates had been washed by sonication at $60^\circ$C with detergent, water, acetone, isopropanol, and then UV-ozone treated) for 5\,s at 400\,rpm followed by 30\,s at 3000\,rpm. The obtained films were annealed on a hotplate at $100^\circ$C for 10\,min. Spin-coating and annealing were carried out in a N$_2$-filled glovebox. The procedure described here is an adaptation of that described in Ref.~\citenum{SI_du2017}.

The lead(II) iodide (PbI$_2$, 99\%) and Phenethylamine ($\geq$99\%) were purchased from Sigma-Aldrich. The hydriodic acid (57\% aqueous solution, stabilized with 1.5\% hypophosphorous acid) was purchased from ABCR. The N,N-Dimethylformamide (99.8\%, Extra Dry over Molecular Sieve, AcroSeal) was purchased from Acros. All chemicals were used as received without further purification. The synthesis of phenethylammonium iodide ((PEA)I) was performed in-house and is described in the Supporting Information, S1.1.

Results of XRD, AFM, and profilometer characterization of the samples are given in Ref.~\cite{SI_Kirstein2022c}. The crystals consist of a stack of about 54 perovskite monolayers, separated from one another by a pair of PEA molecules. The perovskite monolayers are formed by PbI$_4$ corner shared octahedra.
The length of the octrahedra is $2\times 3.17$\,\r{A} \cite{SI_hu2019}, and the pair of PEA molecules separate those monolayers by 10.05\,\r{A} at the closest iodine to iodine contact~\cite{SI_mahal2022}. The unit cell of the layered perovskite structure is larger due to PEA twisting and an octrahedral tilt. The $a$, $b$, $c$ axis lengths reported in Ref.~\citenum{SI_du2017} are $a=b=8.74\,$\r{A}, $c=33.00$\,\r{A} for the room temperature crystal configuration.

\subsection{Synthesis of phenethylammonium iodide (PEA)I}

As a special detail in the description of the sample growth, the synthesis of (PEA)I is presented. 13\,ml of hydriodic acid was added dropwise to a mixture of 10\,ml of phenethylamine and 20\,ml of ethanol (anhydrous) while stirring. The mixture was cooled with an ice bath and stirred for 1\,hour. The resulting solution was evaporated in a rotary evaporator at $50^\circ$C until the liquid was completely removed. The obtained solid was washed three times with diethylether and recrystallized from ethanol with diethylether. The final product of (PEA)I was obtained after drying under vacuum at $50^\circ$C overnight (14\,g). 

\section{Experimental results}

\subsection{Polarization of SFRS}

\label{sec:SI_Polarization}
\begin{figure*}[hbt]
\begin{center}
\includegraphics[width=17.8cm]{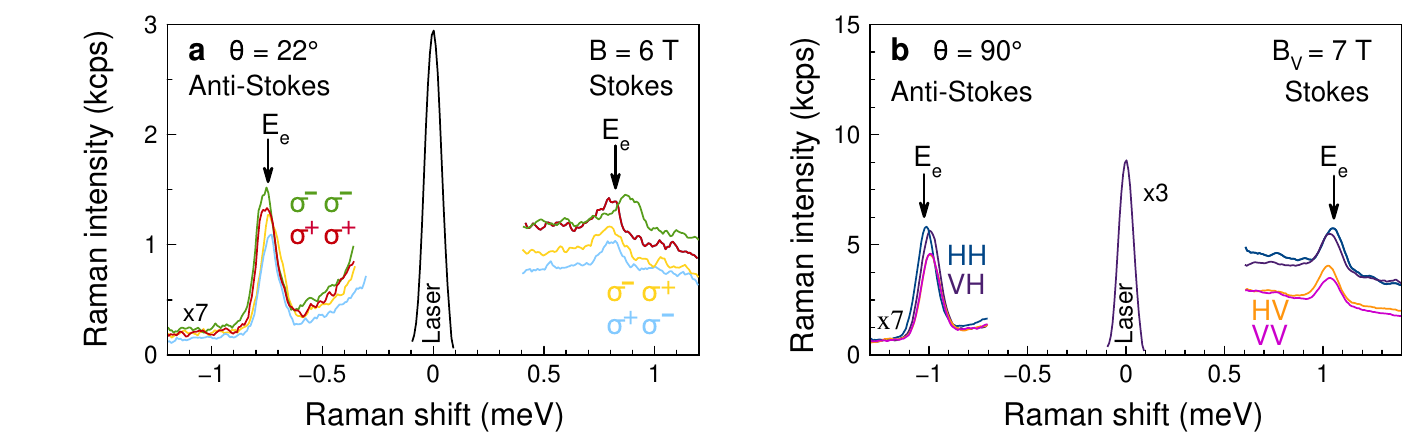}
\caption{
SFRS spectra measured in the anti-Stokes and the Stokes ranges for $E_\text{exc}=2.345$~eV photon energy using a power $P=5.7$~W/cm$^2$ at $T=1.6$~K. 
(a) SFRS in tilted field geometry with $\theta=22^{\circ}$ for both co- and cross-circular polarization at $B=6$~T.
(b) SFRS in Voigt geometry for co- and cross-linear polarization at $B_{\rm V}=7$~T. The linear polarizations $H$ and $V$ are given for excitation / detection. The anti-Stokes spectra are multiplied with a factor of 7 to increase the visibility of the SFRS lines. The electron $E_\text{e}$ and the combined $E_\text{e+h}$ spin-flip Raman peak positions are marked by arrows.
}
\label{fig:polarization}
\end{center}
\end{figure*}

In Figure~\ref{fig:polarization}(a), the SFRS spectra in circularly co-polarized and cross-polarized configuration at $\theta=22^\circ$ for $B=6~$T are shown. 
In the anti-Stokes and Stokes ranges, the electron spin-flip is clearly visible with a Raman shift of $|E_\text{e}| = 0.75~$meV. Note that the absolute SFRS intensities of the electron spin-flip Raman peaks do not differ in the co- and cross-polarization configurations. The minor changes are originating from variations of the resonant photoluminescence background which is weaker in the $\sigma^+/\sigma^+$ configuration (blue spectrum). 

In Figure~\ref{fig:polarization}(b) the SFRS spectra for the different linear co- and cross-polarizations are presented, measured in the Voigt geometry ($\theta=90^\circ$) at $B_{\rm V}=7$~T. The Raman shift of the electron spin-flip is 

$|E_\text{e,(a,b)}|=1.03~$meV. With increasing tilt angle, the spin-flip process efficiency increases, so that the electron spin-flip intensity in the Stokes and anti-Stokes ranges increases approximately five times in comparison to $\theta=22^\circ$.  The comparison of the different linear polarization configurations shows that the absolute electron SFRS peak intensity is only 20\% weaker in the cross-linear configurations than in the co-linear configurations, both for Stokes and anti-Stokes.

\subsection{SFRS temperature dependence}
\label{sec:SI_temperature}

\begin{figure*}[hbt]
\begin{center}
\includegraphics[width=17.8cm]{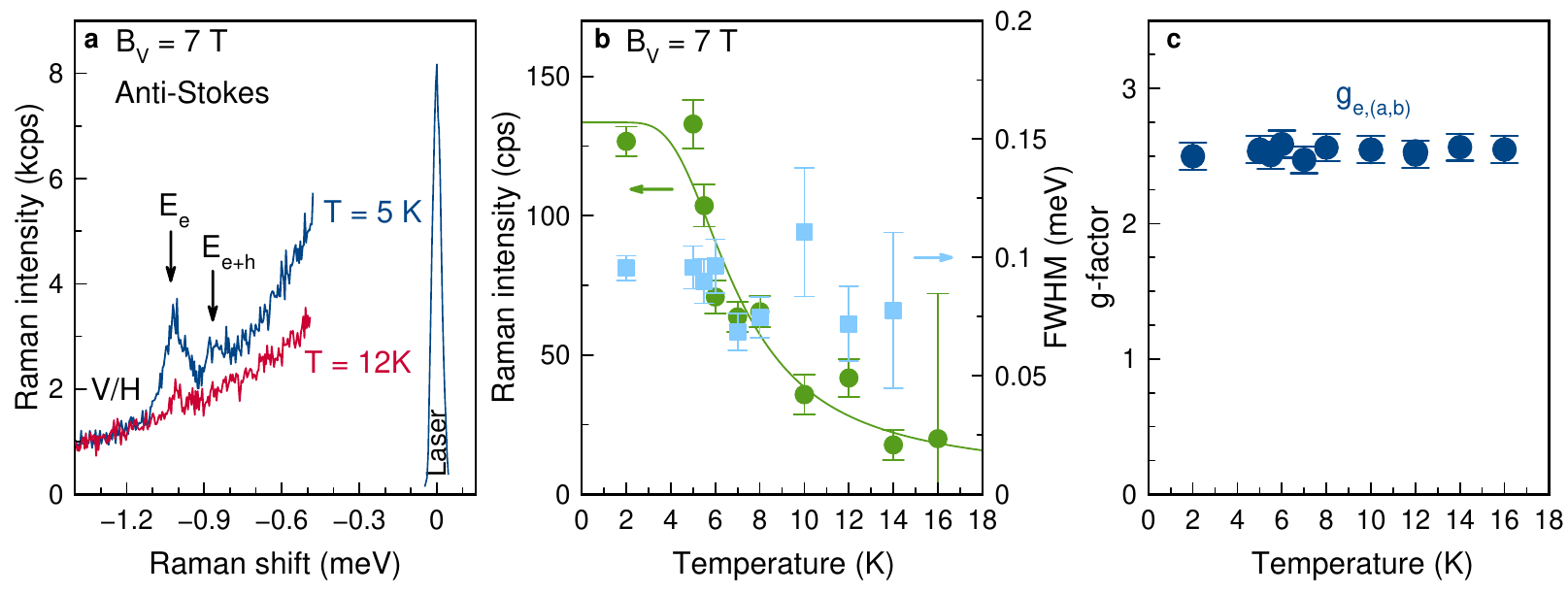}
\caption{Temperature dependence of SFRS in (PEA)$_2$PbI$_4$ measured in the Voigt geometry.
(a)  Anti-Stokes SFRS spectra at $T=5$~K (blue) and $12$~K (red), measured at $B_{\rm V}=7$~T. Both spectra are multiplied by a factor of 10 in order to increase the visibility of the spin-flip signals. 
(b) Temperature dependence of the electron SFRS intensity (green circles) and the full width at half maximum (blue squares). The temperature behavior of the electron intensity follows the Arrhenius-like equation~\eqref{eqn:Arrhenius} shown by the green line fit. 
(c) Temperature  dependence of the electron $g$-factor. 
}
\label{fig:Temperature}
\end{center}
\end{figure*}

Experimental data on the temperature dependence of SFRS measured in the Voigt geometry are given in Figure~\ref{fig:Temperature}. The SFRS anti-Stokes spectra at temperatures of $5$~K and $12$~K are presented in Figure~\ref{fig:Temperature}a. At $T=5$~K the electron ($E_{\rm e}$) and double ($E_{\rm e+h}$) spin-flip lines are pronounced compared to $12$~K, where only the weak electron spin-flip remains visible. The Raman shift is not affected by temperature as shown by the temperature dependence of the electron g-factor in Figure~\ref{fig:Temperature}c: the electron $g$-factor remains constant, $g_\text{e,(a,b)}=2.5\pm0.1$, in the temperature range from 1.6~K up to 16~K, as one would expect from the negligible changes of the band structure in this range. 

The amplitude of the electron spin-flip line decreases with increasing temperature, but is hardly detectable above 16~K. The temperature dependence of the SFRS intensity is shown by the green circles in Figure~\ref{fig:Temperature}b and can be described by the Arrhenius-like equation
\begin{equation}
I(T) = \Big(A \exp{{\Big(\frac{E_\text{A}}{k_B T}}\Big)}+C \Big)^{-1} \, ,
\label{eqn:Arrhenius}
\end{equation}
with the activation energy $E_\text{A}$ and the Boltzmann constant $k_B$. 

From fitting the data, the parameters amount to $E_\text{A}=2.1~$meV, with the amplitude $A = 0.22$ and the constant $C= 0.0075$ (see the line in Figure~\ref{fig:Temperature}b). We suggest that the thermal delocalization of resident electrons is the mechanism that leads to the reduction of the efficiency of the SFRS process. Moreover, in Figure~\ref{fig:Temperature}b the temperature dependence of the full width at half maximum (FWHM) of the electron spin-flip line is presented which has a rather constant value of about 0.1~meV within the examined temperature range.







\end{document}